\documentclass[letter]{aa} 

\usepackage{graphicx}
\usepackage{txfonts}
\usepackage{array}
\newcommand{\teff}{$T_{\rm eff}$}

\def\kms{km~s$^{-1}$}

\def\teff{$T_{\rm eff}$}

\def\vsini{$v\sin i$}
\def\prot{${\rm P}_{rot}$}

\def\wli{${\rm EW}({\rm Li})$}

\def\prot{${\rm P}_{rot}$}

\begin{document}

   \title{The lithium-rotation connection in the newly discovered young stellar stream Psc-Eri (Meingast 1)\thanks{Based on observations collected at the European Organisation for Astronomical Research in the Southern Hemisphere under ESO programme 103.A-9009.} }


\titlerunning{Lithium and rotation in the Psc-Eri stream}
 \authorrunning{Arancibia-Silva et al.}

   \author{J. Arancibia-Silva
          \inst{1,2}
          \and
          J. Bouvier\inst{3}
          \and
          A. Bayo\inst{1,2}
          \and
          P.A.B. Galli\inst{4}
          \and 
          W. Brandner\inst{5}
          \and
          H. Bouy\inst{4}
          \and
          D. Barrado\inst{6}
          }

   \institute{Instituto de F\'{\i}sica y Astronom\'{\i}a, Universidad de Valpara\'{\i}so, Chile\\
         \and
         N\'ucleo Milenio Formaci\'on Planetaria - NPF, Universidad de Valpara\'iso, Av. Gran Breta\~na 1111, Valpara\'iso, Chile\\
         \and
             IPAG, Univ. Grenoble Alpes, 38000 Grenoble,  France\\
             \and
             Laboratoire d'Astrophysique de Bordeaux, Univ. Bordeaux, CNRS, B18N, All\'ee Geoffroy Saint-Hilaire, 33615, Pessac, France\\
             \and 
             Max Planck Institute for Astronomy, Heidelberg, Germany\\
             \and 
             Depto. Astrof\'{\i}sica, Centro de Astrobiolog\'{\i}a (INTA-CSIC), ESAC Campus, Camino Bajo del Castillo s/n, 28692, Villanueva de la Ca\~nada, Spain 
             }

   \date{Accepted}

 
  \abstract
   {As a fragile element, lithium is a sensitive probe of physical processes occurring in stellar interiors.}
   {We aim at investigating the relationship between lithium abundance and rotation rate in low-mass members of the newly discovered 125~Myr-old Psc-Eri stellar stream. }
   {We obtained high resolution optical spectra and measure the equivalent width of the 607.8 nm LiI line for 40 members of the Psc-Eri stream, whose rotational periods have been derived by \cite{Curtis19}.}
   { We show that a tight correlation exists between lithium content and rotation rate among the late-G to early K-type stars of the Psc-Eri stream. Fast rotators are systematically Li-rich, while slow rotators are Li-depleted. This trend mimics the one previously reported for the similar age Pleiades cluster.}
   {The lithium-rotation connection thus seems to be universal over a restricted effective temperature range for low-mass stars at or close to the zero-age main sequence, and does not depend on environmental conditions. }


   \maketitle
%

\section{Introduction}

The evolution of lithium abundance over a star's lifetime reflects transport processes operating in the stellar interior. Lithium, a fragile element that is burned at a temperature of 2.5 MK encountered at the base of the convective zone of low-mass stars, is slowly depleted and its surface abundance steadily decreases over time in solar-type and lower mass stars \citep[see, e.g.,][for a review]{Jeffries14}.

Li depletion is extremely dependent on temperature \citep[e.g,][]{Bildsten97}, making Li abundances a very sensitive diagnostics of the physics of transport processes in stellar interiors. More vigorous transport processes will lead to more severe Li depletion. Beyond the dependence on temperature, Li depletion can also be affected by non-standard transport processes, such as rotational mixing, internal magnetic fields, and by structural changes induced by, for instance, rotation, magnetic activity, metallicity, or accretion \citep[e.g,][]{Pinsonneault90,Zahn92,Ventura98,Piau02,Talon05,Denissenkov10,Eggenberger12b,Theado12}. These additional mechanisms depend at the same time on initial conditions and evolutionary paths. It is therefore important to investigate lithium content in relation to other properties of the stars at different ages and in different environments. 

A connection between lithium abundance and rotation has been reported in young stellar populations at different ages, where rapid rotators are found to have systematically higher Li abundances than slow ones. This relationship is somewhat counter-intuitive as rotational mixing is thought to scale with surface rotation, and would thus predict Li-depleted fast rotators.
\citet{Soderblom93} first reported this trend for the Pleiades (age $\sim$125 Myr), which was recently confirmed and extended by \citet{Bouvier18}. The Li-rotation connection was also investigated at such young ages as $\sim$5 Myr, in the case of NGC 2264 \citep{Bouvier16}, and in the $\beta$ Pic moving group at 20 Myr \citep{Messina16}.
Studying this relationship at different ages and/or in different environments is key to decipher its origin. 

In this paper, we focus on studying this relationship in the newly discovered solar-metallicity ($[Fe/H]=-0.04 \pm\ 0.15)$ stellar stream Psc-Eri \citep{Meingast19}\citep[Meingast 1 in][]{Ratzenbock2020Meingast1PscEri}, which has an age similar to that of the Pleiades \citep{Curtis19}, but is located in an overall very different environment. We combine our new lithium equivalent width (EW(Li)) measurements with the rotational periods reported by \cite{Curtis19} in order to investigate the lithium-rotation connection in this new environment.

\section{Methodology}

\subsection{The stellar sample}

Our starting sample was that of candidate members listed by \cite{Curtis19} who derived rotational periods from TESS light curves \citep[][]{Ricker15}. 
We additionally used \citet{Curtis19} \teff\ estimates, which are indistinguishable from Gaia DR2 values, to select members with \teff\  in the range from 4000 to 5500K, corresponding to spectral type G5-K6. This is the  \teff\ range over which the relationship between lithium and rotation was observed in the Pleiades. We thus obtained a sample of 66 late G and early to mid K-type stars in the Psc-Eri stream. 

\subsection{Observations}

Among this initial sample, we observed 40 stars with the FEROS \citep{Kaufer99} high-resolution spectrograph (R$\sim$48,000) mounted at the ESO/MPG 2.2m telescope at ESO/La Silla Observatory. The observations were obtained from July 24 to August 2, 2019, as part of Programme 103.A-9009. We observed the targets in object-calibration mode which allows for the simultaneous acquisition of the object spectrum and the ThAr lamp. The exposure times ranged from 2 to 25 min depending on the magnitude of the target. We reduced the collected spectra with the FEROS data reduction pipeline which performs bias subtraction, flat-fielding, wavelength calibration, barycentric velocity correction, and merging of the extracted echelle orders into a single one-dimensional spectrum that we used in the forthcoming analysis.  Most spectra have a signal-to-noise ratio (SNR) between 30 and 50. However, a couple of objects have low quality spectra, namely curt\#92 (SNR$\sim$10) and curt\#43 (SNR$\sim$15). Radial velocities (${\rm V}_{r}$) were estimated via cross correlation with a synthetic K0 spectral mask to further correct the spectra and shift it to the rest frame (The values are summarized in Table \ref{res}). Even though, due to visibility / airmass constraints, 40 objects were observed out of the 66 from the previously described sample, this subsample homogeneously covers the range of \teff\ and rotational periods that we wanted to explore.

\subsection{Lithium equivalent width measurements}

EW(Li) were obtained in an automatic fashion described in Appendix A of \cite{Bayo11}. In addition, we independently measured EW(Li) interactively from direct integration and Gaussian fitting using the IRAF/splot command. The comparison of the two methods yielded consistent results for most cases. A few discrepant cases resulted from low SNR spectra (Obj 15), bad pixels affecting the automatic continuum determination (Obj 94), or fast rotation leading to strong line blends (Obj 2, 34) that required a special treatment. 

For the three fast rotators curt\#79, 88, 71 we estimated \vsini\ values of 55, 55, and 61~\kms, respectively, via the previously mentioned K0 mask cross-correlation. The broadening of the lines of these objects lead us to consider the potential impact of line blends on the EW(Li) measurements. We applied the same method as that used for Pleiades fast rotators described in \cite{Bouvier18} (Paper I). Namely, we selected a slow rotator of the same spectral type, produced an artificial Gaussian LiI 6708\AA\ line in its spectrum, and rotationally broadened the spectrum to the \vsini\ of the fast rotator. We then measured EW(Li) in the broadened spectrum and compared it to the input value. As in Paper I, we find that the blends do not affect the EW(Li) measurements for these moderate \vsini\ values. As an additional check, we also used low \vsini\ Psc-Eri members of the same spectral type with varying degree of EW(Li), broadened their spectra, and used them as templates. For instance, Curt\#46 is an intrinsically fast rotator with a period of 1.26d but with moderate \vsini\ (21 \kms), presumably seen at low inclination. It exhibits a deep and relatively narrow lithium line, and is thus a good template to measure EW(Li) in the broadened spectra of other fast rotators. In most cases, this approach yielded similar results as the direct measurements in broadened spectra, which confirms that blends do not significantly affect the derived EW(Li), even for relatively high \vsini\ objects. 

Finally, as in Paper I, we applied a correction on the measured equivalent width to account for the blend by the nearby FeI 6707.441\AA\ line. Following the prescription of \cite{Soderblom93}, we derived the intrinsic (B-V) color of the targets from their Teff, interpolating through \cite{Pecaut13} (B-V)-Teff relationship. Over the (B-V) range of our sample, the FeI blend correction to EW(Li), which is listed in Table~\ref{res}, amounts to 11-19 m\AA, and is thus comparable to our measurement errors. This correction has not been applied to EW(Li) upper limits. Having thus derived EW(Li) for all the objects in our sample, and using rotational periods and \teff\  from  \cite{Curtis19}, we are able to study the relationship between lithium and rotation for the low-mass members of the Psc-Eri stream.

\begin{figure}
	\center
 	\includegraphics[width=0.8\columnwidth]{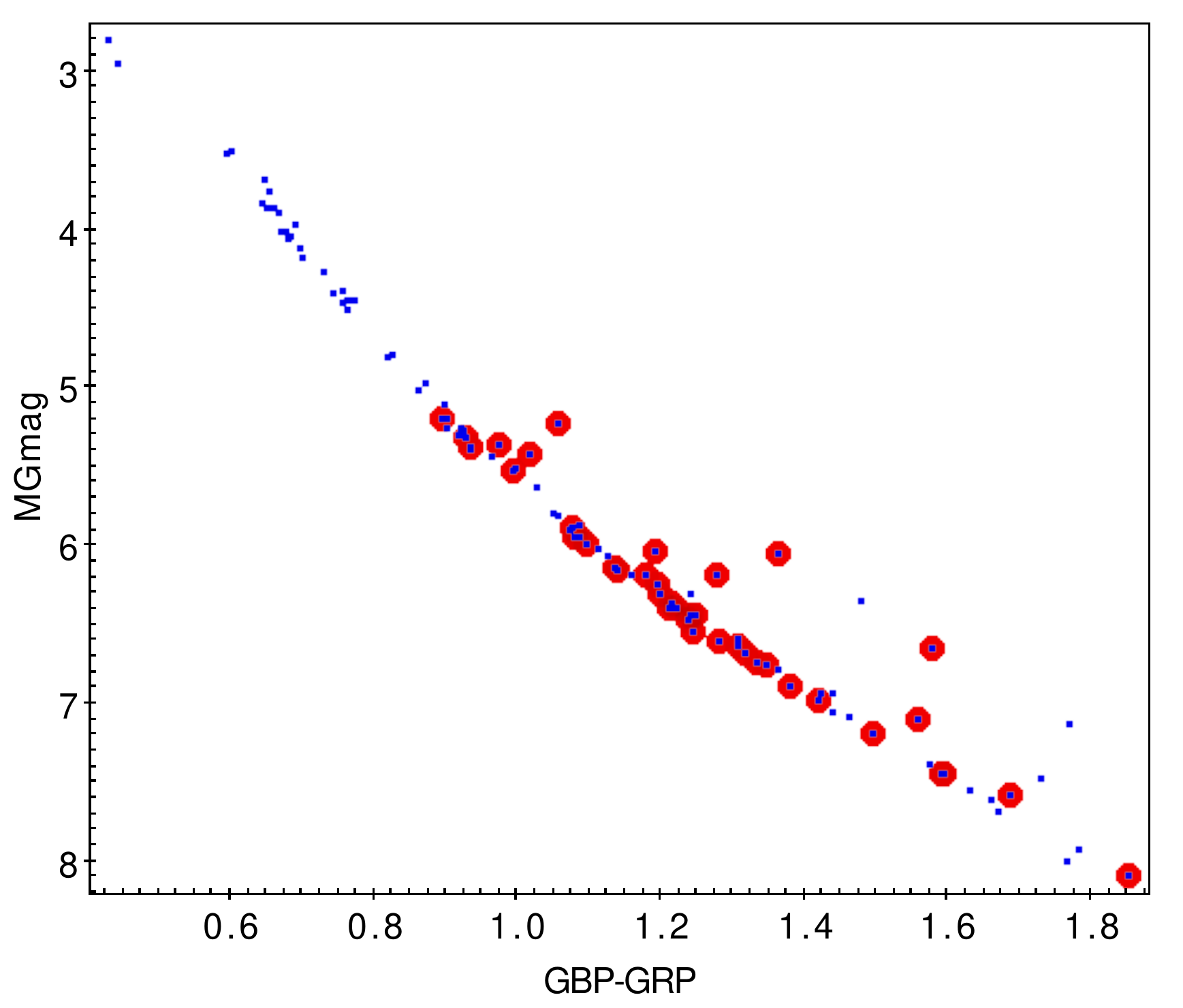}
   \caption{ GAIA's color-magnitude diagram for the Psc-Eri sample. {\it Small blue square:} \cite{Curtis19}'s sample. {\it Large red circles:} Stars with lithium measurements from this study. Note the 7 objects located above the single star main sequence.}
	    \label{cmd}
\end{figure}

\section{Results}
\label{results}

\begin{figure*}
	\includegraphics[width=2.\columnwidth]{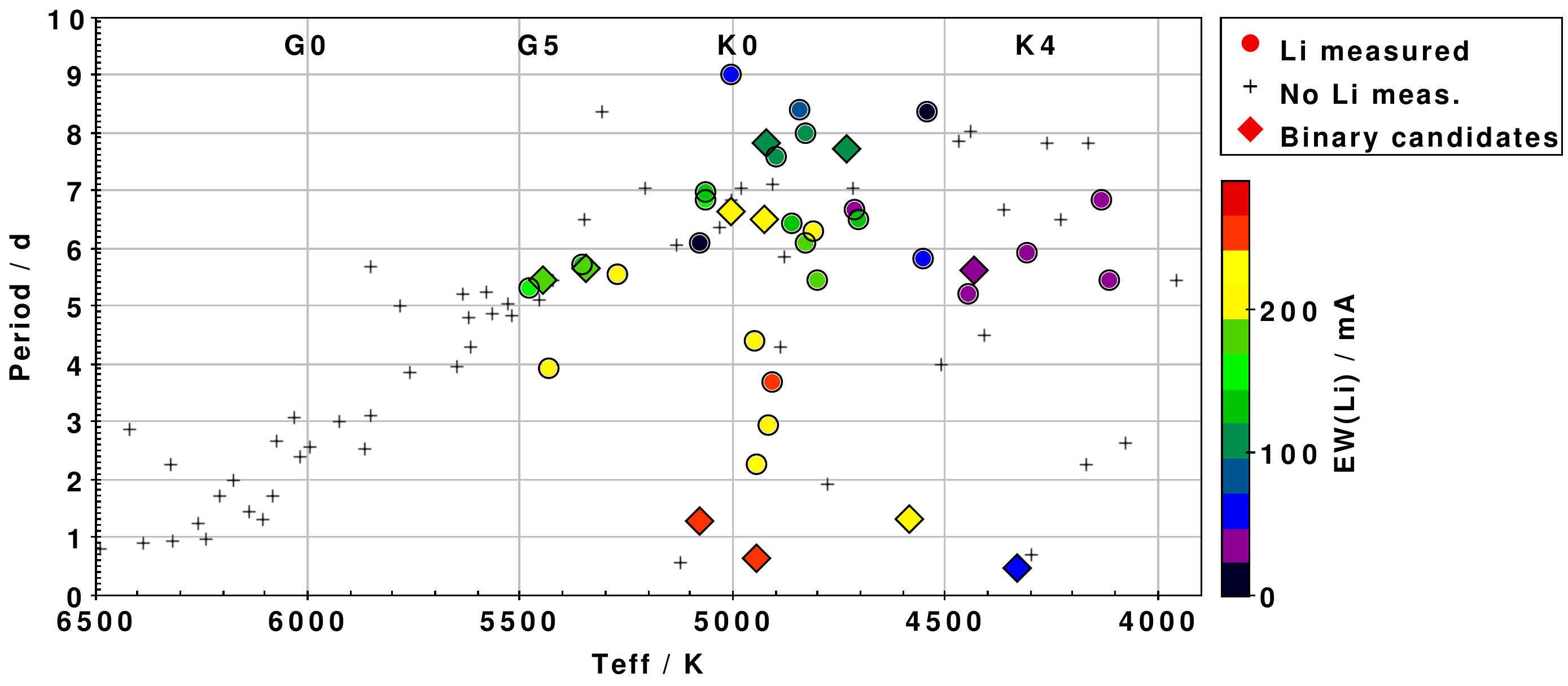}
    \caption{The rotational period distribution of low-mass members of the Psc-Eri stream is plotted as a function of effective temperature. {\it Filled circles:} stars with lithium equivalent width measurements. The color code scales with EW(LiI). {\it Crosses:} Stars without lithium measurement. {\it Filled diamonds:} Bona fide binary systems. A clear trend appears with the fast rotators being systematically more lithium rich than the slow rotators over the \teff\ range 4500-5100~K. }
	    \label{lirot}
\end{figure*}

Before presenting the results of the lithium-rotation relationship for Psc-Eri members below, we discuss a few outliers in the \cite{Curtis19} sample that required further analysis.  

\cite{Curtis19} identified 5 "Slow" stars in their sample, i.e., stars that are lying significantly above the slow rotator sequence of the Psc-Eri stream. Two of those are included in our sample. Curt\#49 is reported to have a 12.22d period. The object has two TESS light curves in the MAST archive from Sectors 4 and 5. The first one is of low quality and does not yield a period. The second one exhibits a smooth modulation and a CLEAN periodogram analysis indicates a clear period of 6.083d. We therefore adopted this value in the following, which brings this object back onto the slow rotator sequence. 
The other "Slow" object, Curt\#92, has a reported period of 11.96d. We queried the MAST archive for this object, but no TESS light curve is provided there. The closest object with a TESS light curve is 397 arcsec away. In any case, the signal to noise ratio of the spectrum we obtained for this object (the poorest quality one in our campaign), SNR$\sim$10, is too low to detect the lithium line in absorption. The uncertainty in the period and poor quality spectrum prompted us to discard this object in the following analysis.

Our preliminary analysis revealed another outlier in the lithium-rotation-\teff\  plane, namely Curt\#75 with a reported period of 1.93 days \citep{Curtis19}, i.e., an intrinsically fast rotator. However, its spectrum exhibits narrow photospheric line profiles and 
no sign of lithium absorption. The observer at the telescope  noticed that two objects were visible on the acquisition camera. Indeed, 2MASS K-band and SDSS2 images reveal an elongated object at this position, most likely a nearly equal-mass binary with a separation of a few arcseconds. With its 21 arcsec-wide pixels, the TESS light curve cannot resolve the system. The FEROS spectrum was obtained for only one of the components and we cannot ascertain that this is the one for which a TESS period has been reported. Indeed, the narrow line profiles seem inconsistent with fast rotation, unless the system is seen at a low inclination. In the light of these uncertainties, we discarded this source from the analysis.

We are thus left with 38 objects with TESS periods reported by \cite{Curtis19} and with EW(Li) measured on the FEROS spectra. The results are listed in Table~\ref{res}. 
Of these, 7 appear to be photometric binaries in the GAIA's color-absolute magnitude diagram shown in Figure~\ref{cmd}. Objects Curt\#46, 79, 90 are well above the single stream sequence, while Curt\#43, 60, 71, 88 are slightly above. Owing to the high accuracy of Gaia data and the narrowness of the single star sequence, we classify the 7 stars as photometric binaries (PB).
In addition to these, to further study possible multiplicity effects, we looked for signs of tight companions, i.e. spectroscopic binaries (SBs). To search for single line spectroscopic binaries (SB1), we compared our ${\rm V}_{r}$ determinations with those of Gaia DR2. Among the objects with good quality spectra, we found five (namely Curt\#36, 52, 57, 66, 79) with differences in ${\rm V}_{r}$ determination above 5$\sigma$. In particular, Curt\#79 and \#71 have been classified by us as PBs and also potential SB1 (although the delta in ${\rm V}_{r}$ for Curt\#71 is only at the 3$\sigma$ level). To complete the spectroscopic binary search, we visually inspected the CCFs looking for signs of resolved spectroscopic binaries (SB2) but we did not find any obvious candidate. With this analysis in hand, the SB binary fraction estimated for this stream is consistent with that determined for loose associations of similar age \citep[$\sim$10\%][]{Elliott14}. 

Regarding multiplicity effects on Li measurements assuming as bona fide binary systems all of PBs and SBs found (Curt\#36, 43, 46, 52, 57, 60, 66, 71, 79, 88, 90), four of our candidate binaries are fast rotators with periods lower than two days, and the other seven are slower rotators, with periods between $\sim$ 5 and $\sim$ 8 days. 
The binary candidates are distributed over nearly the whole range of periods. They also are not uniformly Li-rich or Li-poor, but seem to follow the same tendency as the single stars (see filled diamonds in Figure \ref{lirot}). This suggest that the trend that we are reporting in the Lithium content is most probably tied to the rotation of the stars and not to their multiple nature.

\begin{table*}
\caption{Stellar sample properties from \cite{Curtis19} and lithium equivalent width measurements}             
\label{res}      
\centering                          
\tabcolsep=0.15cm
\begin{small}
\begin{tabular}{l l l l l l l l l l l l l l}        
\hline\hline                 
GaiaDR2 & Curt & RA(2000) & DEC(2000) & \teff & G & $M_{G}$ & G$_{bp}$-G$_{rp}$ & \prot & EW(FeI) & \wli & rms & ${\rm V}_{r}$ & $v\sin i$ \\
&&hh:mm:ss & dd:mm:ss & K & mag & mag & mag & d & m\AA & m\AA & m\AA & km/s & km/s \\
\hline
2349094158814399104 & 79* & 00:47:18.0 & -22:45:08.1 & 4585 & 10.929 & 6.051 & 6.051 & 1.3 & 17 & 202 & 10 & 29.36 & 55 \\
4975223840046231424 & 81 & 00:47:38.5 & -47:41:45.8 & 4551 & 11.514 & 6.985 & 6.985 & 5.8 &  & $<$50 &  & 15.67 & 7 \\
2355466790769878400 & 96 & 00:55:21.5 & -21:24:03.7 & 4135 & 12.341 & 7.578 & 7.578 & 6.84 &  & $<$30 &  & 17.63 & $<$6 \\
5029398079322118912 & 52* & 01:13:42.4 & -31:11:39.6 & 5003 & 10.804 & 5.959 & 5.959 & 6.64 & 14 & 196 & 5 & 20.24 & $<$6 \\
4984094970441940864 & 101 & 01:21:49.7 & -42:01:22.3 & 4113 & 12.731 & 8.099 & 8.099 & 5.45 &  & $<$30 &  & 20.61 & 7 \\
2484875735945832704 & 65 & 01:24:24.7 & -03:16:39.0 & 4842 & 11.791 & 6.398 & 6.398 & 8.4 & 15 & 82 & 13 & 16.84 & 7 \\
2491594263092190464 & 56 & 02:10:22.3 & -03:50:56.7 & 4944 & 11.533 & 6.153 & 6.153 & 2.26 & 14 & 225 & 10 & 19.37 & 19 \\
5117016378528360448 & 91 & 02:17:14.6 & -27:16:41.9 & 4308 & 12.551 & 7.451 & 7.451 & 5.92 &  & $<$30 &  & 21.65 & 8 \\
5118895478259982336 & 90* & 02:26:07.0 & -24:54:49.0 & 4431 & 11.728 & 6.658 & 6.658 & 5.61 & 18 & 31 & 5 & 21.83 & 10 \\
2488721720245150336 & 87 & 02:26:53.3 & -05:17:45.2 & 4444 & 12.485 & 7.198 & 7.198 & 5.2 &  & $<$30 &  & 21.19 & 8 \\
5129876953722430208 & 76 & 02:29:28.5 & -20:12:16.8 & 4829 & 11.695 & 6.739 & 6.739 & 8.0 & 15 & 99 & 5 & 22.7 & 7 \\
2496200774431287424 & 36* & 02:30:58.8 & -03:03:04.9 & 5448 & 10.415 & 5.328 & 5.328 & 5.45 & 11 & 173 & 5 & 22.25 & $<$6 \\
5179037454333642240 & 64 & 02:39:10.9 & -05:32:22.5 & 4859 & 11.765 & 6.373 & 6.373 & 6.42 & 15 & 125 & 5 & 19.23 & 8 \\
5161117923061794688 & 74 & 02:59:52.0 & -09:47:35.8 & 4800 & 12.063 & 6.639 & 6.639 & 5.45 & 15 & 184 & 10 & 20.44 & 8 \\
5045955865443216640 & 40 & 03:00:46.9 & -37:08:01.5 & 5434 & 10.323 & 5.363 & 5.363 & 3.9 & 11 & 198 & 10 & 21.91 & $<$6 \\
5159567164990031360 & 88* & 03:04:46.0 & -12:16:57.9 & 4330 & 12.613 & 7.098 & 7.098 & 0.45 & 19 & 50 & 30 & 19.27 & 55 \\
7324465427953664 & 61 & 03:05:14.1 & +06:08:53.5 & 4947 & 12.043 & 6.247 & 6.247 & 4.4 & 14 & 195 & 10 & 18.61 & 10 \\
5103353606523787008 & 46* & 03:18:03.8 & -19:44:14.2 & 5077 & 10.473 & 5.227 & 5.227 & 1.26 & 13 & 242 & 5 & 20.41 & 25 \\
5106733402188456320 & 71* & 03:24:25.2 & -15:50:05.4 & 4944 & 11.517 & 6.197 & 6.197 & 0.62 & 14 & 245 & 10 & 40.06 & 61 \\
5168681021169216896 & 29 & 03:29:30.3 & -07:10:13.8 & 5478 & 10.827 & 5.204 & 5.204 & 5.32 & 11 & 168 & 5 & 20.78 & 8 \\
3247412647814482816 & 80 & 03:32:30.9 & -06:13:09.1 & 4714 & 12.327 & 6.898 & 6.898 & 6.66 & 16 & 34 & 10 & 22.87 & 8 \\
5114686272872474880 & 69 & 03:47:25.8 & -12:32:30.9 & 5006 & 12.634 & 6.55 & 6.55 & 9.0 & 14 & 61 & 15 & 20.15 & 7 \\
4842810376267950464 & 38 & 03:47:56.3 & -41:56:24.9 & 5355 & 10.762 & 5.383 & 5.383 & 5.7 & 12 & 143 & 5 & 18.75 & $<$6 \\
3243665031151732864 & 50 & 03:48:38.3 & -06:41:52.6 & 5063 & 11.46 & 5.953 & 5.953 & 6.84 & 13 & 151 & 5 & 22.04 & $<$6 \\
3245140743257978496 & 43* & 03:54:01.0 & -06:14:14.6 & 5344 & 11.146 & 5.434 & 5.434 & 5.66 & 12 & 193 & 15 & 21.7 & 8 \\
3193528950192619648 & 41 & 03:57:04.0 & -10:14:00.9 & 5270 & 11.297 & 5.534 & 5.534 & 5.54 & 12 & 195 & 5 & 20.88 & $<$6 \\
5083255496041631616 & 49 & 03:57:35.1 & -24:28:42.2 & 5078 & 11.131 & 5.885 & 5.885 & 6.08 & 13 & 7 & 5 & 22.98 & $<$6 \\
5097262136011410944 & 62 & 03:58:54.7 & -17:05:53.2 & 4899 & 11.649 & 6.311 & 6.311 & 7.58 & 14 & 105 & 5 & 22.45 & 7 \\
5096891158212909312 & 59 & 04:12:46.0 & -16:19:29.1 & 4907 & 11.994 & 6.187 & 6.187 & 3.68 & 14 & 255 & 10 & 21.25 & 13 \\
4871041608622321664 & 60* & 04:28:28.9 & -33:53:45.1 & 4924 & 11.63 & 6.037 & 6.037 & 6.5 & 14 & 215 & 10 & 21.21 & 7 \\
2594993646533642496 & 67 & 22:31:13.9 & -17:04:52.4 & 4829 & 11.934 & 6.451 & 6.451 & 6.1 & 15 & 174 & 10 & 9.96 & $<$6 \\
2402197409339616768 & 57* & 22:39:01.4 & -18:52:55.7 & 4921 & 11.219 & 6.155 & 6.155 & 7.8 & 14 & 115 & 5 & 13.09 & $<$6 \\
2596395760081700608 & 53 & 22:39:53.5 & -16:36:23.3 & 5065 & 11.508 & 5.994 & 5.994 & 6.97 & 13 & 126 & 5 & 10.75 & 7 \\
2433715455609798784 & 70 & 23:36:52.1 & -11:25:01.7 & 4810 & 11.737 & 6.442 & 6.442 & 6.3 & 15 & 214 & 10 & 14.66 & 8 \\
2393862836322877952 & 66* & 23:40:37.5 & -18:11:37.9 & 4734 & 11.485 & 6.472 & 6.472 & 7.7 & 16 & 99 & 10 & 16.88 & 7 \\
2390974419276875776 & 72 & 23:48:32.4 & -18:32:57.4 & 4706 & 11.583 & 6.618 & 6.618 & 6.5 & 16 & 123 & 10 & 17.14 & 8 \\
2418664520110763520 & 63 & 23:49:55.1 & -15:43:42.0 & 4917 & 11.607 & 6.396 & 6.396 & 2.94 & 14 & 215 & 10 & 16.87 & 15 \\
2339984636258635136 & 77 & 23:56:53.7 & -23:17:24.6 & 4545 & 11.475 & 6.765 & 6.765 & 8.35 & 17 & 22 & 5 & 15.8 & 7 \\
\hline
\end{tabular}
\begin{flushleft}
* Bona fide binary systems (see Section \ref{results} for details.)
\end{flushleft}
\end{small}
\end{table*}

Figure~\ref{lirot} illustrates the resulting relationship between lithium content and rotation among the late G and early K-type stars of the Psc-Eri stream. As previously reported by \cite{Curtis19}, stars in this mass range exhibit a wide range of rotational periods. The same behavior is seen in the Pleiades cluster, and this is precisely this similarity that prompted \cite{Curtis19} to assign the same age of 125 Myr to the Psc-Eri stream. The lithium measurements reported here also show a wide dispersion over this \teff\  range, from undetectable ($\leq$30 m\AA) up to 260 m\AA. Assuming the same age and metallicity for the Psc-Eri stream as for the Pleiades, the latter value correspond to a nearly pristine lithium abundance of A[Li]$\simeq$2.8 \citep[see][]{Bouvier18}. 

Figure~\ref{lirot} very clearly reveals the dependence of lithium content on rotation rate at a given \teff, over the \teff\  range from 4500 to 5100~K. There, fast rotators with periods shorter than 4d have EW(Li) ranging from 200 to 255~m\AA, while slow rotators with periods longer than 6d exhibit lithium equivalent widths less than 215~m\AA\ down to undetectable levels ($\leq$30 m\AA), with a median value of 110~m\AA. As we previously found in the Pleiades cluster, there seems to be a tight lithium-rotation relationship valid for all stars in this \teff\  range on the zero-age main sequence, with slow rotators being lithium depleted compared to fast ones. Whether this relationship extends over a wider range of effective temperature awaits additional lithium measurements for both hotter and cooler stars of the stream.  

\section{Discussion and Conclusions}

\cite{Meingast19} reported a new, nearby stellar stream identified from the 6D position-velocity data provided by Gaia DR2. They estimated an age of 1 Gyr based on the location of member giant stars relative to model isochrones and derived a solar-metallicity by  cross-correlating their sample with LAMOST DR4.  \cite{Curtis19} derived rotational periods from TESS light curves for 101 low-mass members of the stream, that they named Psc-Eri. From the striking similarity of the rotation period versus mass diagrams of the Psc-Eri stream and of the Pleiades cluster, they assigned an age of 125 Myr to the former. Indeed, the distribution of rotation rates as a function of mass is quite sensitive to age, especially close to the zero-age main sequence \citep[e.g.,][]{Bouvier14, Gallet15}. 

The connection we report here between lithium and rotation for low-mass stars of the Psc-Eri stream is also strikingly similar to that reported earlier for the Pleiades cluster (Paper I). Fast rotators are systematically Li-rich compared to slow ones, a relationship valid both for single and binary systems over a limited \teff\  range. This result when first reported by \cite{Soderblom93} was quite unexpected. Most stellar evolution models would predict that faster rotators experience enhanced internal mixing, and would thus burn lithium more efficiently \citep[e.g.,][]{Pinsonneault89}. Fast rotators ought to be lithium depleted while the opposite was observed. Subsequent studies including the present one unambiguously point to a higher lithium content in fast rotators at or close to the zero-age main sequence. In Paper I, we summarized 3 classes of models that attempt to explain this paradoxical result, and we will only briefly recall them here. The reader is referred to \citet{Bouvier16} for a more extensive discussion of their success and limitations. 

\cite{Bouvier08} related the lithium-rotation connection observed on the ZAMS to the pre-main sequence rotational history of young solar-type stars. Current angular momentum evolution models predict that slow rotators remain locked to their circumstellar disk over a longer duration than fast ones \citep[e.g.,][]{Gallet15}. As a result, a larger amount of internal differential rotation develops in slow rotators, which triggers enhanced mixing and leads to PMS Li depletion, while fast rotators maintain their original Li content \citep{Eggenberger12b}. Alternatively, \cite{Somers14, Somers15} suggested that radius inflation in fast, magnetically active stars reduces PMS Li burning due to a lower temperature at the base of the convective zone. Assuming a relationship between rotation and magnetic activity during the PMS, the slow, more moderately active stars would not experience radius inflation, and would thus start to deplete lithium earlier. Finally, \cite{Baraffe17} argued that stellar rotation has a direct impact on the efficiency of internal transport processes, and proposed that fast rotation reduces the penetration of convective plumes into the radiative core, thus leading to a lower depletion rate in fast rotators. Any of these models, or combination of them, has the potential to explain the lithium-rotation connection reported here and in previous studies. Further studies will be needed to attempt to discriminate between them. 

The clear lithium-rotation relationship we report here for the Psc-Eri low-mass members has two main implications. Firstly, the lithium-rotation connection seems to be universal for low-mass stars in a specific \teff\  or mass range at or close to the zero-age main sequence. It is reported here for the ZAMS stream Psc-Eri and has been previously reported in various star forming regions, moving groups, and young open clusters with ages between 5 and 125~Myr \citep{Messina16, Bouvier16, Bouvier18}. Secondly, this relationship does not seem to be affected by environmental conditions. While the Psc-Eri stream and Pleiades cluster probably have similar ages \citep[][Barrado et al., in prep.]{Curtis19}, they have quite different kinematic properties. The former is in the process of dissolving into the galactic field, while the latter is still a well defined young open cluster. Presumably, the Pleiades cluster formed from a very rich and dense environment, similar to the Orion Nebula Cluster \citep{Kroupa01}, while the Psc-Eri stream may have originated from a rich albeit lower density OB or T association at birth. These different initial conditions have apparently had no impact on the lithium-rotation connection, whose origin is thus to be found in the physics of pre-main sequence stellar evolution. 

Further characterization of the lithium-rotation connection for stellar populations of different ages, metallicities, and environments will help to guide the development of pre-main sequence stellar evolution models able to account for this result.

\begin{acknowledgements}
AB and JB acknowledge funding from the ECOS 170006 - Conicyt C17U01 project for mutual visits. Most of this study was completed during a stay at University Valparaiso in the framework of this program. JB gratefully acknowledges the hospitality of IfA members during his visit. This research has received funding from the European Research Council (ERC) under the European Union's Horizon 2020 research and innovation programme (grant agreements No 742095 SPIDI: Star-Planets-Inner Disk-Interactions, http://spidi-eu.org; and No 682903, P.I. H. Bouy, COSMIC DANCE, http://astrophy.u-bordeaux.fr/en/cosmic-dance/), and from the French State in the framework of the "Investments for the future" Program, IdEx Bordeaux, reference ANR-10-IDEX-03-02. This research has been partially funded by the Spanish State Research Agency (AEI) Projects No.ESP2017-87676-C5-1-R and No. MDM-2017-0737 Unidad de Excelencia “María de Maeztu”- Centro de Astrobiología (INTA-CSIC).
\end{acknowledgements}

\bibliographystyle{aa} 
\bibliography{arancibia} 

\end{document}